 \documentstyle[preprint,epsf,eqsecnum,aps]{revtex}
\def\D0{\mbox{D\O }}
\def\Et{$E_{T}$}

\def\MEt{\mbox{$E\kern-0.57em\raise0.19ex\hbox{/}_{T}$}\ }
\def\MEtp{\mbox{$E\kern-0.57em\raise0.19ex\hbox{/}_{T}$}}
\def\Fe{$\varepsilon$}

\begin{document}

\title{Search for the Trilepton Signature from the
Associated Production of SUSY $\tilde{\chi}_{1}^{\pm} 
\tilde{\chi}_{2}^{0}$ Gauginos}

%
\author{                                                                      
B.~Abbott,$^{28}$                                                             
M.~Abolins,$^{25}$                                                            
B.S.~Acharya,$^{43}$                                                          
I.~Adam,$^{12}$                                                               
D.L.~Adams,$^{37}$                                                            
M.~Adams,$^{17}$                                                              
S.~Ahn,$^{14}$                                                                
H.~Aihara,$^{22}$                                                             
G.A.~Alves,$^{10}$                                                            
E.~Amidi,$^{29}$
% ...
\\                                                                            
\vskip 0.50cm                                                                 
\centerline{(D\O\ Collaboration)}                                             
\vskip 0.50cm                                                                 
}                                                                             
\address{                                                                     
\centerline{$^{1}$Universidad de los Andes, Bogot\'{a}, Colombia}             
\centerline{$^{2}$University of Arizona, Tucson, Arizona 85721}               
\centerline{$^{3}$Boston University, Boston, Massachusetts 02215}             
\centerline{$^{4}$Brookhaven National Laboratory, Upton, New York 11973}      
\centerline{$^{5}$Brown University, Providence, Rhode Island 02912}           
\centerline{$^{6}$Universidad de Buenos Aires, Buenos Aires, Argentina}       
\centerline{$^{7}$University of California, Davis, California 95616}          
\centerline{$^{8}$University of California, Irvine, California 92697}         
\centerline{$^{9}$University of California, Riverside, California 92521}      
\centerline{$^{10}$LAFEX, Centro Brasileiro de Pesquisas F{\'\i}sicas,        
                  Rio de Janeiro, Brazil}                                     
\centerline{$^{11}$CINVESTAV, Mexico City, Mexico}                            
\centerline{$^{12}$Columbia University, New York, New York 10027}             
\centerline{$^{13}$Delhi University, Delhi, India 110007}                     
\centerline{$^{14}$Fermi National Accelerator Laboratory, Batavia,            
                   Illinois 60510}                                            
\centerline{$^{15}$Florida State University, Tallahassee, Florida 32306}      
\centerline{$^{16}$University of Hawaii, Honolulu, Hawaii 96822}              
\centerline{$^{17}$University of Illinois at Chicago, Chicago,                
                   Illinois 60607}                                            
\centerline{$^{18}$Indiana University, Bloomington, Indiana 47405}            
\centerline{$^{19}$Iowa State University, Ames, Iowa 50011}                   
\centerline{$^{20}$Korea University, Seoul, Korea}                            
\centerline{$^{21}$Kyungsung University, Pusan, Korea}                        
\centerline{$^{22}$Lawrence Berkeley National Laboratory and University of    
                   California, Berkeley, California 94720}                    
\centerline{$^{23}$University of Maryland, College Park, Maryland 20742}      
\centerline{$^{24}$University of Michigan, Ann Arbor, Michigan 48109}         
\centerline{$^{25}$Michigan State University, East Lansing, Michigan 48824}   
\centerline{$^{26}$Moscow State University, Moscow, Russia}                   
\centerline{$^{27}$University of Nebraska, Lincoln, Nebraska 68588}           
\centerline{$^{28}$New York University, New York, New York 10003}             
\centerline{$^{29}$Northeastern University, Boston, Massachusetts 02115}      
\centerline{$^{30}$Northern Illinois University, DeKalb, Illinois 60115}      
\centerline{$^{31}$Northwestern University, Evanston, Illinois 60208}         
\centerline{$^{32}$University of Notre Dame, Notre Dame, Indiana 46556}       
\centerline{$^{33}$University of Oklahoma, Norman, Oklahoma 73019}            
\centerline{$^{34}$University of Panjab, Chandigarh 16-00-14, India}          
\centerline{$^{35}$Institute for High Energy Physics, 142-284 Protvino,       
                   Russia}                                                    
\centerline{$^{36}$Purdue University, West Lafayette, Indiana 47907}          
\centerline{$^{37}$Rice University, Houston, Texas 77005}                     
\centerline{$^{38}$Universidade Estadual do Rio de Janeiro, Brazil}           
\centerline{$^{39}$University of Rochester, Rochester, New York 14627}        
\centerline{$^{40}$CEA, DAPNIA/Service de Physique des Particules,            
                   CE-SACLAY, Gif-sur-Yvette, France}                         
\centerline{$^{41}$Seoul National University, Seoul, Korea}                   
\centerline{$^{42}$State University of New York, Stony Brook,                 
                   New York 11794}                                            
\centerline{$^{43}$Tata Institute of Fundamental Research,                    
                   Colaba, Mumbai 400005, India}                              
\centerline{$^{44}$University of Texas, Arlington, Texas 76019}               
\centerline{$^{45}$Texas A\&M University, College Station, Texas 77843}       
}                                                                             

\maketitle
\begin{abstract}
We report on a search for the trilepton signature from the
associated production of supersymmetric gaugino pairs,
$\tilde{\chi}_{1}^{\pm} \tilde{\chi}_{2}^{0}$, within the context of
minimal supersymmetric models that conserve $R$-parity.
This search uses 95 pb$^{-1}$ of data taken with the \D0 detector at 
Fermilab's Tevatron
collider at $\sqrt{s} = 1.8$ TeV. 
No evidence of a trilepton signature has been found, and a limit 
on the production cross section times branching fraction to trileptons
as a function of $\tilde{\chi}_{1}^{\pm}$ mass is given. 
\end{abstract}


The Standard Model (SM) is very successful, but
there are a number of theoretical arguments that suggest it will
break down at the TeV scale unless it is extended. 
One argument involves the necessity of fine-tuning the parameters 
of the Higgs scalar potential in order to obtain a Higgs
mass near the electro-weak scale.
Supersymmetry (SUSY) is among the leading possibilities
for an extension of the SM.
SUSY relates bosons to fermions and introduces for every SM
particle a supersymmetric partner
that differs in spin by 1/2.
The SUSY electro-weak gauge particles
(gauginos) are mixtures of the SUSY partners of the $W$, $Z$,
$\gamma$ and Higgs bosons. The charged and neutral gauginos are 
denoted by $\tilde{\chi_{i}}^{\pm}\, \{i=1,2\}$ and 
$\tilde{\chi_{i}}^{0}\, \{i=1,2,3,4\}$.
We consider only minimal supergravity (SUGRA)~\cite{susy1} models 
or minimal unified scale (GUT)~\cite{susy2} inspired models
that are $R$-parity conserving.
$R$-parity conservation requires that SUSY particles be produced in pairs 
and that the lightest SUSY particle (LSP) be absolutely stable. 
In the models
we investigate, this LSP is the $\tilde{\chi}_{1}^{0}$,
which is a candidate for cold dark matter.

This letter describes a search for the production of
$\tilde{\chi}_{2}^{0} \tilde{\chi}_{1}^{\pm}$ pairs which
decay producing three isolated charged leptons 
plus missing transverse energy (\MEtp)~\cite{theory}. 
The $\tilde{\chi}_{2}^{0}$ in this case decays into
two charged leptons plus a LSP, and the $\tilde{\chi}_{1}^{\pm}$ decays into
a charged lepton, a neutrino, and a LSP. The backgrounds
to this hadronically quiet trilepton signature are small. 
The two highest transverse energy (\Et) 
leptons have moderate to high \Et\ ($>$15 GeV), while the lepton with the third 
highest \Et\ can be rather soft.
Even though
two LSP's and a neutrino contribute to the \MEtp, 
the angular correlation between these particles
is weak resulting in moderate \MEtp.
The $\tilde{\chi}_{2}^{0} \tilde{\chi}_{1}^{\pm}$
production cross section times branching fraction to trileptons
($\sigma\times B(3\ell)\,$)
varies greatly as a function of the model parameters.

We search for evidence of $\tilde{\chi}_{2}^{0} \tilde{\chi}_{1}^{\pm}$
production in four channels
containing electrons ($e$) and muons ($\mu$):
$eee$, $ee\mu$, $e\mu\mu$, $\mu\mu\mu$. Due to detection inefficiencies, 
we ignore taus and
their leptonic decay products.
The integrated luminosities 
for the search
in the above four channels are respectively 94.9 pb$^{-1}$,
94.9 pb$^{-1}$, 89.5 pb$^{-1}$, and 75.3 pb$^{-1}$, obtained
during the 1994--1995 Tevatron collider run at $\sqrt{s} = 1.8$ TeV.
Previous searches~\cite{oldprl}~\cite{cdf} at the Tevatron 
for trilepton signatures were conducted using the
1992--1993 collider data. 

The \D0 detector is described in detail elsewhere~\cite{det}.
It consists of central tracking chambers
without a magnetic field, a finely segmented, hermetic 
uranium/liquid-argon 
sampling calorimeter, and a muon spectrometer. 
Electrons are measured with
an energy resolution of 15\%/$\sqrt{E}$, and the muon momentum
is measured with a resolution expressed as
$\sigma (1/p) = 0.18(p-2)/p^{2} \oplus 0.003$ ($E$, $p$ measured in GeV
and GeV/c).

The event selection is optimized based on signal and background Monte Carlo
simulations and background data. We require three isolated leptons
satisfying standard \D0 identification 
requirements~\cite{id}.
Electrons are required to satisfy the isolation requirement 
${\cal I} < 0.1$,
where ${\cal I} = (E_{\mbox{\scriptsize TOT}} 
- E_{\mbox{\scriptsize EM}})/E_{\mbox{\scriptsize EM}}$.
$E_{\mbox{\scriptsize TOT}}$ is the energy of the electron candidate within 
the electromagnetic
(EM) and hadronic portions of the calorimeter that is within a cone of radius 
$R = \sqrt{\Delta\eta^{2}+\Delta\phi^{2}} = 0.4$, where
$\eta$ is the pseudorapidity and $\phi$ is the azimuthal angle.
$E_{\mbox{\scriptsize EM}}$ is the energy of the candidate electron within the
EM calorimeter and first layer of the hadronic calorimeter
that is within a cone of $R = 0.2$. Muons are required
not to have any reconstructed jet (\Et\ $>$ 8 GeV) within $R = 0.5$.
The minimum lepton \Et\ is 5 GeV; however,
one or two of the leptons are required to be 2 GeV above 
the trigger thresholds. The triggers used in this analysis are 
listed in Table~\ref{tab:trig}.

Muons are required to have
$\mid\eta\mid$ $<$ 1.0. To eliminate instrumental
backgrounds with large mismeasured \MEt due to tails in the muon momentum
distribution, $\Delta \phi$ between
the highest \Et\ muon and the \MEt must be $< (\pi - 0.1)$ radians
and any signature muon must have $\Delta \phi > 0.1$ relative to 
the \MEtp. 
To reject the
cosmic ray background, we require that any two signature muons 
have $\Delta \phi_{\mu \mu} < (\pi - 0.1)$ radians. 
This dimuon back-to-back cut
also helps to eliminate a significant portion of the  
$Z/\gamma^{*}$ boson to dimuon
background.

We also require signature specific
cuts in the four channels. 
For the $eee$ channel, we 
require the
$e\mbox{\MEtp}$ or $2e\mbox{\MEtp}$ trigger, 
and we require \MEt $>$ 15 GeV. 
We exclude events with an
invariant mass from the two highest \Et\ electrons in the 
mass range of 81 to 101 GeV/c$^{2}$, and
the two highest \Et\ electrons must 
have $\Delta \phi_{ee} < (\pi - 0.2)$ radians. 
These cuts eliminate the main background of
$Z/\gamma^{*}$ bosons with an additional ``electron''
(denoted as \Fe)
originating from a jet which fluctuated into an EM cluster or 
from a converted photon which produced two unresolved electrons. 
The effect on the background
of altering the cuts on the \MEt and the \Et\ of the third most 
energetic electron
as estimated from Monte Carlo studies is given in Table~\ref{tab:one}. 
Also given is the actual
number of events seen. The data agree well with
the $Z/\gamma^{*}$ boson background estimates. 
With the cuts of 5 GeV on the third electron and \MEt $>$ 
15 GeV, we see no events in the $eee$ channel
with an expected background of 0.34$\pm$0.07 events.

For the $ee\mu$ channel we require the $e\mbox{\MEtp}$,
$2e\mbox{\MEtp}$, or
$e\mu$ trigger. We also require \MEt $>$ 10 GeV.
With these selections and the above generic
requirements
on electrons and muons, 
we see no events. We expect 0.61$\pm$0.36
background events from three main sources: 
$Z/\gamma^{*}\rightarrow\tau\tau +$ \Fe, semi-leptonic decays of
heavy ($b$ or $c$) quark pairs $+$ \Fe, and
$Z/\gamma^{*}\rightarrow e e + \mu$ from a heavy quark decay. 
Mass and angle cuts are not made on the electrons in this channel, since
the rate of $Z/\gamma^{*}\rightarrow e e + \mu$ events is smaller by 
about an order of magnitude than the rate of
$Z/\gamma^{*}$ + \Fe\ background events in the $eee$ channel. 
Relaxing the isolation requirement on the muon and the
10 GeV \MEt requirement allows one 
event to pass
with an expected background of 1.41$\pm$0.67 events.
This event has a dielectron mass of 90 GeV/c$^{2}$ indicating that it
is a $Z\rightarrow e e +\mu$ candidate, which is consistent with
this background source being the largest contributor
to the total background for these cuts.

The major backgrounds for the $e\mu\mu$ channel are 
$Z/\gamma^{*}$ bosons $+$ \Fe, 
$J/\psi$ $+$ \Fe, 
and heavy quark
pairs $+$ \Fe. For this channel, we require the $\mu$, 
$2\mu$,
or $e\mu$ trigger. 
To reject low mass dimuon
events (e.g., $J/\psi$), we require that the dimuon 
invariant mass be greater
than 5 GeV/c$^{2}$.
We also require \MEt $>$ 10 GeV.
With this selection we see
no candidate events with an expected background of 0.11$\pm$0.04 events.
To verify our background estimate, we relax the electron identification
requirements, thereby dramatically increasing the number of events with
misidentified electrons.
We see 31 events with 27.3$\pm$5.5 events expected.
About 80\% of this is estimated to come from heavy quark pairs.

$Z/\gamma^{*}$ bosons and heavy quark pairs are the major contributors of 
background to 
the $\mu\mu\mu$ channel. In this channel we require the $\mu$
or $2\mu$ trigger. The dimuon invariant mass for any 
two of the three muons must be greater than 5 GeV/c$^{2}$.
We also require \MEt $>$ 10 GeV.
We see no events with this selection. We expect 0.20$\pm$0.04
background events. Without the \MEt cut we find one event with an
expected background of 0.75$\pm$0.27 events. We interpret this event
to be consistent with a heavy quark pair with \MEt $=$ 1.3 GeV.

The channel specific
selection requirements 
are summarized
in Table~\ref{tab:cuts}. A summary of the total backgrounds expected for 
our final event selection
and the integrated
luminosity is given in Table~\ref{tab:back}.
The luminosities vary from channel to channel due to different prescales 
for the various triggers.

The signal efficiencies are derived from {\footnotesize ISAJET}~\cite{isajet}
Monte Carlo 
processed with a {\footnotesize GEANT}~\cite{geant} 
simulation of the \D0 detector
and a simulation of the \D0 trigger. 
The model parameters for our full simulation
were chosen to give $M_{\tilde{\chi}_{1}^{\pm}} =
M_{\tilde{\chi}_{2}^{0}}$ within 1 GeV and $M_{\tilde{\chi}_{2}^{0}} = 
2M_{\tilde{\chi}_{1}^{0}}$ within 10\%,
since these relationships hold approximately for many choices of parameters 
in SUGRA models.  We generate events 
in the four signatures with $\tilde{\chi}_{1}^{\pm}$ masses between 45
and 124 GeV/c$^{2}$. The efficiency ranges from 1.6\% at 45 GeV/c$^{2}$
to 11.1\% at 124 GeV/c$^{2}$ for the $eee$ channel and decreases as the
signature includes more muons down to the range of 0.54\% to 2.17\% for 
the $\mu\mu\mu$ channel. The efficiencies for the channels with
muons are smaller due to the reduced $\eta$ acceptance of muons compared to
electrons and the lower identification efficiency for muons.

These efficiencies and our resulting limit on $\sigma\times B(3\ell)$ are 
applicable to many choices of SUSY
model parameters.
To estimate under what conditions our signal efficiencies apply, 
we have studied the {\footnotesize ISAJET} particle spectra from
$\tilde{\chi}_{1}^{\pm}$ $\tilde{\chi}_{2}^{0}$
production for a large number of choices (scenarios) of
the five SUGRA model parameters. These parameters and the
chosen range of values are: the common scalar mass at
the unified (GUT) scale, $1 \leq m_0 \leq 100$ GeV/c$^{2}$;
the common fermion mass at the GUT scale, 
$60 \leq m_{1/2} \leq 155$ GeV/c$^{2}$; the ratio of the vacuum
expectation values of the two Higgs doublets at the electroweak scale, 
$1.5 \leq \tan \beta \leq 6$;
the soft trilinear SUSY breaking parameter at the GUT scale, 
$-200 \leq A_0 \leq 200$; and
the sign of the Higgsino mass term, $\mu$.  

We find that 99\% of the scenarios studied with
$M_{\tilde{\chi}_{2}^{0}}/M_{\tilde{\chi}_{1}^{0}} \geq 1.8$,
$M_{\tilde{\chi}_{2}^{0}} - M_{\tilde{\chi}_{1}^{\pm}} 
\geq -1.0$ GeV/c$^2$, and $M_{\tilde{\chi}_{1}^{\pm}} > 45$ GeV/c$^2$
have efficiencies that are $\geq$ 0.9 times the efficiency for the
case where
$M_{\tilde{\chi}_{1}^{\pm}} =  
M_{\tilde{\chi}_{2}^{0}}
= 2M_{\tilde{\chi}_{1}^{0}}$. However, if the
masses of the SUSY partners of the charged leptons, $\tilde{l}$,
are lighter than one or both of $\tilde{\chi}_{2}^{0}$
and $\tilde{\chi}_{1}^{\pm}$, in order for our efficiencies
to be applicable, $M_{\tilde{\chi}_{2}^{0}} - M_{\tilde{l}} > 7.0$ GeV,
$M_{\tilde{\chi}_{1}^{\pm}} - M_{\tilde{l}} > 7.0$ GeV, and
$M_{\tilde{l}} - M_{\tilde{\chi}_{1}^{0}} > 15.0$ GeV.

\begin{figure}
\epsfxsize=6.0in
\epsfysize=6.0in
\epsffile{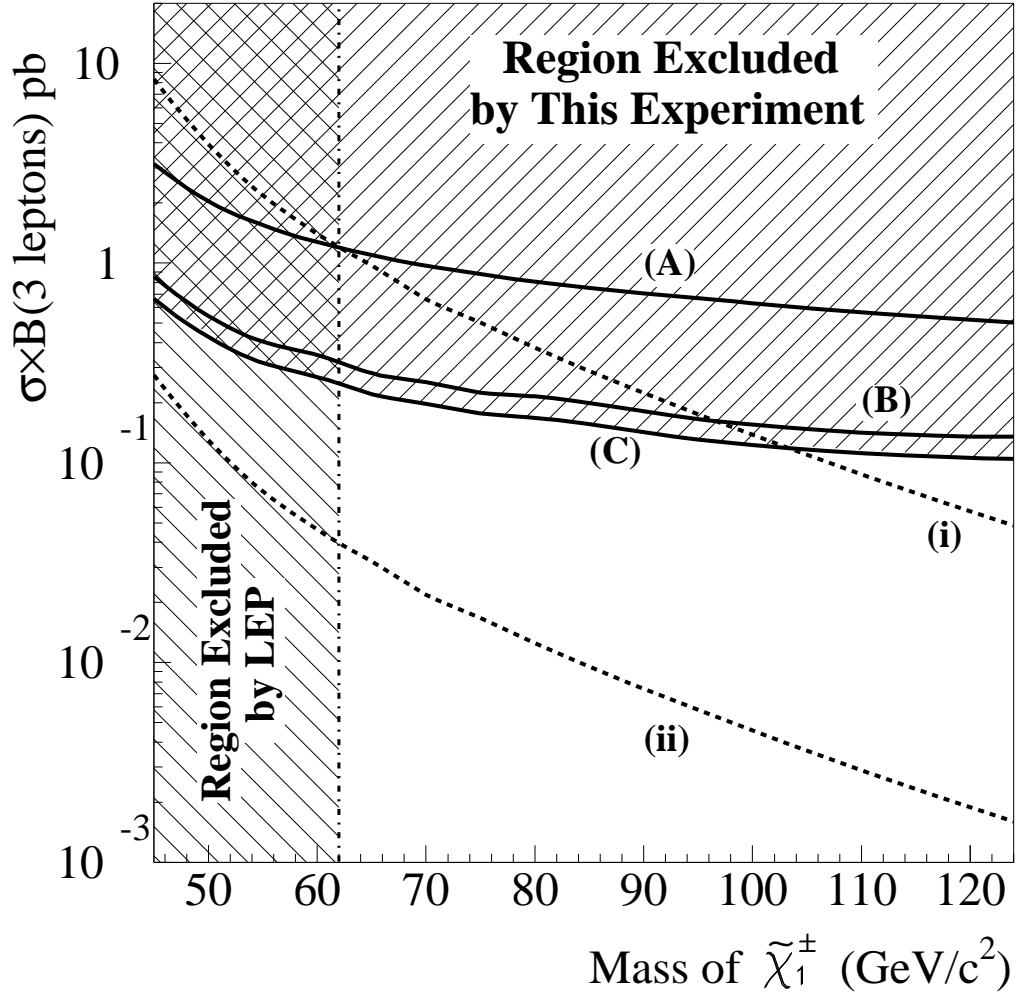}
\caption{The 95\% CL upper limit on $\sigma\times B(3\ell)$
versus $\tilde{\chi}_{1}^{\pm}$ mass for any given channel. 
(A): limit from 1992--1993 data, (B): limit from 1994--1995 data,
(C): combined limit, (i) and (ii): theoretical $\sigma\times B(3\ell)$.}
   \label{fig:limit}
\end{figure}

Combining all four channels and assuming that the branching fractions
for the decay of $\tilde{\chi}_{1}^{\pm} \tilde{\chi}_{2}^{0}$ 
to the four channels are equal, we calculate 
the 95\% CL upper limit~\cite{method} on $\sigma\times B(3\ell)$
for any one channel.
This limit takes into account the total statistical and systematic 
uncertainties of the
analysis. These total uncertainties range from 10\% for the $eee$ channel
to 20\% for the $\mu\mu\mu$ channel. The previously published limit
(based on 12.5 pb$^{-1}$ of 1992--1993 data)~\cite{oldprl} as a function of 
$\tilde{\chi}_{1}^{\pm}$ mass is given 
in Fig.~\ref{fig:limit}
as the top solid curve (A). 
The limit from the 1994--1995 data is shown 
as the middle solid
curve (B), and the limit from the combined data set is given 
as the lower solid curve (C). We exclude the region above this curve.
The combined limit ranges from 0.66 pb
at $M_{\tilde{\chi}_{1}^{\pm}} = 45$ GeV/c$^{2}$ 
down to 0.10 pb at $M_{\tilde{\chi}_{1}^{\pm}} = 124$ GeV/c$^{2}$.
The top dashed curve (i) is the theoretical cross section (GUT inspired) for 
$\tilde{\chi}_{1}^{\pm} \tilde{\chi}_{2}^{0}$ production
from {\footnotesize ISAJET}
times the maximum branching fraction to trileptons of 1/9 for any one channel. 
This represents the maximum possible $\sigma\times B(3\ell)$ 
for one exclusive trilepton channel.
The bottom dashed
curve (ii) is the total cross section times the product
of 
the SM branching fractions of 
$W$ and $Z$
bosons to any one generation of charged lepton (0.0036).
This is given to illustrate the typical variation of 
$\sigma\times B(3\ell)$ within SUSY models, but 
in some scenarios the branching fraction can approach zero.
Also given as the shaded region to the left is the 95\% CL
lower limit of 62 GeV/c$^{2}$
on the $\tilde{\chi}_{1}^{\pm}$ mass 
from the OPAL $\sqrt{s} = 161$ GeV data for the conditions on the SUSY
model parameters as given in Ref.~\cite{lep}. Other limits from the
LEP 130 GeV and 136 GeV data are discussed in Ref.~\cite{lep2}.
$\tilde{\chi}_{1}^{\pm}$ masses below 45 GeV/c$^{2}$ have been 
excluded by previous searches at LEP~\cite{lep3}.

In conclusion we find no evidence of 
$\tilde{\chi}_{1}^{\pm} \tilde{\chi}_{2}^{0}$ production in the current 
\D0 data set. We have set
a 95\% CL upper limit on $\sigma\times B(3\ell)$ to any one
channel as a function of $\tilde{\chi}_{1}^{\pm}$ mass.

%
We thank the staffs at Fermilab and collaborating institutions for their
contributions to this work, and acknowledge support from the 
Department of Energy and National Science Foundation (U.S.A.),  
Commissariat  \` a L'Energie Atomique (France), 
State Committee for Science and Technology and Ministry for Atomic 
   Energy (Russia),
CNPq (Brazil),
Departments of Atomic Energy and Science and Education (India),
Colciencias (Colombia),
CONACyT (Mexico),
Ministry of Education and KOSEF (Korea),
CONICET and UBACyT (Argentina),
and the A.P. Sloan Foundation.

\begin{table} 
\caption{Triggers used in SUSY gaugino search.}
\label{tab:trig}
\begin{tabular}{c c}
Trigger & Requirements \\
\hline
$e\mbox{\MEt}$ & $\geq 1e,\mbox{\Et} > 20$ GeV and \MEt $>$ 15 GeV \\
\hline
$2e\mbox{\MEt}$ & $\geq 1e,\mbox{\Et} > 12$ GeV 
and $\geq 1e,\mbox{\Et} > 7$ GeV \\
 & and \MEt $>$ 7 GeV \\
\hline
$e\mu$     & $\geq 1e,\mbox{\Et} > 7$ GeV and 
$\geq 1\mu,\mbox{\Et} > 8$ GeV \\
\hline
$\mu$          & $\geq 1\mu,\mbox{\Et} > 15$ GeV \\
\hline
$2\mu$   & $\geq 2\mu,\mbox{\Et} > 3$ GeV \\
\end{tabular}
\end{table}

\begin{table} 
\caption{The number of events observed and background estimates 
for $Z/\gamma^{*} \rightarrow ee +$ \Fe \
for various \MEt and third electron \Et\ (denoted as $E_{T}^{3}$)
cuts in the $eee$ selection.}
\label{tab:one}
\begin{tabular}{ c c c c c}
\mbox{}&\multicolumn{2}{c}{\MEt $>$ 10 GeV}&
\multicolumn{2}{c}{\MEt $>$ 15 GeV}\\
\mbox{}& \#expected & \#seen & \#expected & \#seen\\
$E_{T}^{3}$ (GeV) & \multicolumn{4}{c}{\mbox}\\
\hline
2 & 4.8$\pm$0.7 & 5 & 1.8$\pm$0.3 & 2\\
\hline
3 & 2.3$\pm$0.4 & 1 & 0.88$\pm$0.17 & 0 \\
\hline
4 & 1.3$\pm$0.2 & 0 & 0.49$\pm$0.10 & 0 \\
\hline
5 & 0.9$\pm$0.2 & 0 & 0.34$\pm$0.07 & 0 \\
\end{tabular}
\end{table}

\begin{table}        
\caption{Summary of cuts used in SUSY gaugino search.}
\label{tab:cuts}
\begin{tabular}{c c c c c}
Channel & $eee$ & $ee\mu$ & $e\mu\mu$ & 
$\mu\mu\mu$ \\
\hline
Trigger & $e\mbox{\MEtp}$, $2e\mbox{\MEt}$ & $e\mbox{\MEtp}$, 
$2e\mbox{\MEtp}$, $e\mu$ & $e\mu$, $\mu$, $2\mu$ & $\mu$, $2\mu$ \\
\hline
Mass cut & $\mid M_{ee} - M_{Z^{0}}\mid > 10$ GeV/c$^{2}$ &
-- & $M_{\mu\mu} > 5$ GeV/c$^{2}$ & 
$M_{\mu\mu} > 5$ GeV/c$^{2}$ \\
 & & & & (all 3 $\mu$ pairs) \\
\hline
\MEt  & $>$ 15 GeV & $>$ 10 GeV & $>$ 10 GeV &
$>$ 10 GeV \\
\hline
angle cuts & $\mid \pi - \Delta\phi_{ee}\mid >0.2$ & -- &
$\mid \pi - \Delta\phi_{\mu\mu}\mid >0.1$ & 
$\mid \pi - \Delta\phi_{\mu\mu}\mid >0.1$ \\
   & (two highest \Et\ electrons) &  &  & (all 3 $\mu$ pairs)\\
\hline
Lepton \Et & 
\multicolumn{4}{c}{For all channels, 2 GeV above trigger 
for one or two leptons} \\
 & \multicolumn{4}{c}{and for all three leptons 
\Et $>$ 5 GeV } \\
\end{tabular}
\end{table}

\begin{table} 
\caption{Summary of expected backgrounds and integrated luminosity. No
events were seen in any channel.}
\label{tab:back}
\begin{tabular}{c c c c c}
 Channel   & $eee$ & $ee\mu$ & $e\mu\mu$ & $\mu\mu\mu$ \\
 Luminosity & 94.9 pb$^{-1}$ & 94.9 pb$^{-1}$ & 
89.5 pb$^{-1}$ & 75.3 pb$^{-1}$\\
 Background & 0.34$\pm$0.07 & 0.61$\pm$0.36 & 0.11$\pm$0.04 & 0.20$\pm$0.04\\
\end{tabular}
\end{table}

\end{document}